\documentclass[final]{svjour2}
\usepackage{graphicx}
\usepackage{rotating}
\usepackage{amssymb}
\usepackage{mathptmx}
\makeatletter
\journalname{Journal of Low-Temperature Physics}

\newcommand{\w}{\omega}
\newcommand{\Log}{\ell n}
\newcommand{\TK}{T_{\rm K}}

\newcommand{\Tc}{T_{\rm c}}

\newcommand{\afs}{AF$_{\rm S}$}
\newcommand{\afl}{AF$_{\rm L}$}

\newcommand{\CeAu}{CeCu$_{6-x}$Au$_x$}

\newcommand{\YRS}{YbRh$_2$Si$_2$}

\newcommand{\CaSr}{Ca$_{2-x}$Sr$_x$RuO$_4$}

\begin{document}

\newcommand{\hdblarrow}{H\makebox[0.9ex][l]{$\downdownarrows$}-}

\title{Orbital-selective Mott transitions:\\Heavy fermions and beyond}

\author{Matthias Vojta}

\institute{Institut f\"ur Theoretische Physik, Universit\"at zu K\"oln, 50937 K\"oln\\
\email{vojta@thp.uni-koeln.de}}

\date{June 8, 2010}

\maketitle



\begin{abstract}
Quantum phase transitions in metals are often accompanied by violations of Fermi liquid
behavior in the quantum critical regime. Particularly fascinating are transitions beyond
the Landau-Ginzburg-Wilson concept of a local order parameter. The breakdown of the
Kondo effect in heavy-fermion metals constitutes a prime example of such a transition. Here, the
strongly correlated $f$ electrons become localized and disappear from the Fermi surface,
implying that the transition is equivalent to an orbital-selective Mott transition, as
has been discussed for multi-band transition-metal oxides. In this article, available
theoretical descriptions for orbital-selective Mott transitions will be reviewed, with an
emphasis on conceptual aspects like the distinction between different low-temperature phases and
the structure of the global phase diagram. Selected results for quantum critical
properties will be listed as well. Finally, a brief overview is given on experiments
which have been interpreted in terms of orbital-selective Mott physics.
%
\end{abstract}


\section{Introduction}

The Fermi-liquid phenomenology is central to our understanding of the low-tem\-per\-a\-ture
behavior of metals \cite{bape91,hvl}.
Therefore, it is of equally fundamental interest to study metallic
states which apparently violate Fermi-liquid behavior. Conceptually, non-Fermi liquid
behavior at low temperatures can occur either in (meta)stable non-Fermi liquid phases or
in the quantum critical regime of zero-tem\-per\-ature phase transitions.\footnote{
Technically, both
cases correspond to physics controlled by either infrared stable or unstable
renormalization-group fixed points.}
While our knowledge about concrete non-Fermi liquid {\em phases} in spatial dimensions
$d\geq 2$ is rather limited (with one class -- an orbital-selective Mott phase --
emerging in the context of this article), non-Fermi liquid behavior arising from
quantum criticality has been studied intensively over the last years. Starting with the
pioneering work of Hertz \cite{hertz}, a lot of work has been invested into quantum phase
transitions (QPT) with conventional (local) order parameters, most prominently
ferromagnetic and antiferromagnetic transitions in metals. In these cases, a bosonic
order parameter field carries the critical dynamics and is coupled to low-energy
particle-hole pairs. Assuming that this coupling can be absorbed by appropriate
modifications of the bosonic action (such as Landau damping of order-parameter
fluctuations), one arrives at the so-called Landau-Ginzburg-Wilson (LGW) (or Hertz)
description of metallic QPT \cite{hertz,millis,moriya}.
A large body of theoretical results have been obtained in this framework \cite{hvl}.
Experimentally, antiferromagnetic QPT in some metals match the LGW predictions, while in
others (mainly of the heavy-fermion class) more singular behavior has been detected
\cite{hvl,geg_rev}. This has motivated alternative ideas, in particular the study of phase
transitions beyond local order parameters, but with topological character. Here, much
less results are available, but it can be expected that this broad class of phase
transitions may hold the explanation to various puzzles in correlated condensed matter
physics.

In this article, I will review a particular type of such QPT --
orbital-selective (OS) Mott transitions - which has emerged in distinct areas of correlation
electron physics. On the one hand, the breakdown of the Kondo effect in heavy-fermion
metals has been extensively discussed
\cite{schroeder98,schroeder00,coleman01,si01,si03,flst1,flst2},
triggered by a growing body of experimental data on heavy-fermion quantum critical points (QCP)
which have been found to be inconsistent with the LGW predictions.
Consequently, it has been assumed that the heavy quasiparticles disintegrate and the entire
Fermi surface collapses at the QCP -- this exciting scenario constitutes in a sense the most
drastic violation of the assumptions of the LGW theory.
Instead, the criticality is carried by emergent degrees of freedom
associated with the Kondo effect.
On the other hand, for correlated multi-orbital transition-metal oxides, the possibility of
partial Mott transitions has been discussed. Here, the first proposal was made by
Anisimov {\em et al.} \cite{anisimov02} for \CaSr\ in order to explain the coexistence of
spin-1/2 moments and metallicity at $x=0.5$ \cite{naka00}.
On the theory side, both flavors of transitions have been investigated
using variants of dynamical mean-field theory (DMFT), and effective field-theory
descriptions have been put forward to capture the low-temperature critical behavior.
As will be discussed below, the Kondo-breakdown and orbital-selective Mott quantum phase
transitions are conceptually identical, i.e., the Kondo breakdown can be regarded as a
special limit of an orbital-selective Mott transition. Therefore, the term ``OS Mott
transition'' will be used throughout this article in the heavy-fermion context as well.

In the phase with partial Mott localization, equivalently the absence of Kondo screening
(short: ``OS Mott phase''),
the fate of the localized spin moments in the limit $T\to 0$ requires special
attention. These moments may eventually order magnetically, or they might form a paramagnetic
spin liquid due to magnetic frustration or strong quantum fluctuations. In the latter
case, a true metallic non-Fermi liquid phase can emerge, dubbed fractionalized Fermi
liquid FL$^\ast$ in Ref.~\cite{flst1}. If magnetic frustration can be tuned separately
from OS Mott physics, then various phases appear possible. In fact, recent experimental
progress has prompted discussions about the structure of a ``global'' phase diagram, to
be described below.

A particularly interesting aspect of orbital-selective Mott transitions is that they are
accompanied by critical Fermi surfaces \cite{senthil08}. Current theoretical ideas imply that the
fermionic spectrum becomes critical over an entire sheet of the Fermi surface, with the
corresponding quasiparticle weight vanishing at the transition.
This also implies a complete reconstruction of Fermi-surface across the QPT,
leading to strong signatures in low-temperature transport experiments.

\subsection{Outline}

The body of this article is organized as follows:
Sec.~\ref{sec:breakdown} is devoted to the phenomenology of quantum phase transitions in
heavy-fermion metals. We shall discuss a variety of low-temperature phases with and
without symmetry breaking, together with possible transitions between them, leading to
the construction of ``global'' phase diagrams. We shall also outline some of the
microscopic approaches to Kondo-breakdown QPT which have been used to obtain concrete results.
Sec.~\ref{sec:osmott} will discuss theoretical studies of orbital-selective Mott
transitions in correlated multi-band Hubbard models.
Sec.~\ref{sec:flst} gives a more detailed account on the metallic spin-liquid phase which
may emerge as a result of a OS Mott (or Kondo-breakdown) transition.
In Sec.~\ref{sec:FS} we briefly discuss the fascinating aspect of critical Fermi
surfaces, which accompany such QPT.
Sec.~\ref{sec:obs} focuses on the experimental detection of OS Mott transitions.
Selected results shall be listed for thermodynamic and transport properties of both the
quantum critical regime and the OS Mott phase. Special emphasis is put on the question
how an OS Mott transition may be distinguished from other quantum phase transitions.
Finally, Sec.~\ref{sec:exp} contains an overview on materials for which the existence of
OS Mott transitions has been discussed in the literature.
An outlook will close the article.


\section{Breakdown of the Kondo effect in heavy-fermion metals}
\label{sec:breakdown}

Consider a Kondo lattice model in dimensions $d\geq 2$,
with a unit cell containing one $c$ and $f$ orbital each,
\begin{equation}
\label{eq:KLM}
\mathcal{H}_{\rm KLM}=\sum_{{\vec k}\sigma} (\epsilon_{\vec k}-\mu) c_{{\vec k}
\sigma}^{\dagger}c_{{\vec k} \sigma} +
J \sum_{i} {\vec S}_i \cdot {\vec s}_i,
\end{equation}
where the chemical potential $\mu$ controls the filling $n_c$ of the
conduction ($c$) band with dispersion $\epsilon_{\vec k}$,
and ${\vec s}_{i}= \sum_{\sigma \sigma'}
c_{i\sigma}^{\dagger} {\vec \tau}_{\sigma \sigma'} c_{i\sigma'} / 2$
is the conduction electron spin density on site $i$.
The conduction band filling will be denoted by $n_c$, with $n_c = L^{-d} \sum_{{\vec k}\sigma}
\langle c_{{\vec k} \sigma}^{\dagger}c_{{\vec k} \sigma} \rangle$ where $L^d$ is the number
of lattice sites.
Sometimes it is useful to explicitly include a Heisenberg-type exchange
interaction between the $f$ electron local moments ${\vec S}_i$,
${\cal H}_{\rm I} = \sum_{ij} I_{ij} {\vec S}_i \cdot {\vec S}_j$,
which may originate from superexchange (or RKKY) interactions.\footnote{
While the Kondo lattice model, Eq.~(\ref{eq:KLM}), contains
the physics of the indirect RKKY interaction between local
moments, it is convenient for the theoretical discussion
to introduce an (additional) explicit inter-moment interaction.
}

\begin{figure}
\begin{center}
\includegraphics[%
  width=0.9\linewidth,
  keepaspectratio]{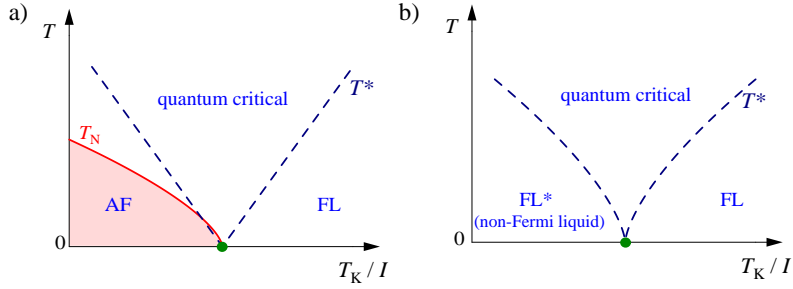}
\end{center}
\caption{ (Color online)
Schematic phase diagram of the Kondo lattice described by
Eq.~(\ref{eq:KLM}), as function of temperature and the ratio between the strength of Kondo
screening, $\TK$, and inter-moment exchange $I$.
a) Conventional scenario due to Doniach \cite{doniach77}, where antiferromagnetic (AF) long-range order
is realized for $\TK\ll I$.
b) If magnetic order is suppressed due to geometric frustration or strong quantum fluctuations of the local
moments, then a topological phase transition towards a non-Fermi liquid phase FL$^\ast$
may occur \cite{flst1}. The dashed crossover lines bound the quantum critical regime. Note that in
both phase diagrams, additional crossover lines are present, influencing both
thermodynamic and transport properties.
}
\label{fig:doniach}
\end{figure}

The model in Eq.~(\ref{eq:KLM}) is believed to capture most aspects of the physics of
heavy-fermion metals. In particular, it is accepted that phase transitions can
be driven by a competition of the lattice Kondo effect, which quenches the local moments
and favors a paramagnetic ground state, and the magnetic interaction between the local $f$
moments \cite{doniach77}. These two competing effects may be characterized by their
respective energy scales, the Kondo temperature $\TK$ (with $\Log\TK\propto -1/J$)
and the typical exchange energy $I$. Varying their ratio leads to phase diagrams as shown
in Fig.~\ref{fig:doniach} which will be discussed in the following.

\subsection{Low-temperature phases and Fermi volume}
\label{sec:fvol}

The heavy Fermi liquid (FL), realized for $\TK\gg I$, is characterized by Kondo screening
of the local moments ${\vec S}_i$. As a result, the local-moment electrons contribute to
the Fermi volume, such that the Fermi volume is ``large'',
\begin{equation}
{\cal V}_{\rm FL} = K_d (n_{\rm tot}\,{\rm mod}\,2)
\end{equation}
with $n_{\rm tot}=n_c+n_f = n_c+1$, in agreement with Luttinger's theorem \cite{oshi2}.
Here, ${\rm mod}\,2$ implies that full bands are not counted, $K_d = (2\pi)^d/(2 v_0)$ is
a phase space factor, with $v_0$ the unit cell volume, and the factor of 2 accounts for
the spin degeneracy of the bands. In this FL phase, the $f$ electrons are usually termed
``delocalized''.

For $\TK\ll I$, the most natural assumption is that magnetic order is realized. For
generic band filling, two distinct types of magnetically ordered metals appear possible.
(i) The magnetism can arise from a spin-density wave instability of the heavy FL state.
Here, Kondo screening is essentially intact, with a weak polarization of the $f$
electrons which are still ``delocalized''. We will refer to such a state as SDW
metal.
(ii) Deep in the ordered phase one might think about a different kind of magnetic metal,
with no Kondo screening and well-established local moments which order due to exchange
interactions \cite{yama}. This state, with ``localized'' $f$ electrons, may be called a
local-moment magnetic (LMM) metal.
Both phases can be expected to have Fermi-liquid properties.

We shall now discuss the Fermi volume for the simplest case of collinear commensurate
antiferromagnetism with an even number $N$ of sites in the AF unit cell.
Here, the spin degeneracy of the bands is preserved.
(The same considerations apply to other orders which enlarge the unit cell and preserve
the spin degeneracy, like valence-bond-solid order.)
The existence of lattice symmetry breaking implies a ``backfolding'' of the bands
into the reduced Brillouin zone.
For the SDW metal, this results in a Fermi volume
\begin{equation}
{\cal V}_{\rm SDW} = K'_d (N n_{\rm tot}\,{\rm mod}\,2)
\end{equation}
with $K'_d = K_d/N$. Remarkably, this value of ${\cal V}_{\rm SDW}$ equals the Fermi
volume of the LMM metal,
\begin{equation}
{\cal V}_{\rm LMM} = K'_d (N n_c\,{\rm mod}\,2).
\end{equation}
Hence, there is no sharp distinction in the Fermi volume between the two cases of
``itinerant'' and ``localized'' $f$ electrons in the presence of antiferromagnetism
(assuming an even number of $f$ electrons per unit cell). We note that a distinction
between the SDW and LMM metals may exist regarding the Fermi surface topology, see
Sec.~\ref{sec:af_kondo}.

In the context of Kondo-breakdown transitions, an unconventional type of phase has been
first proposed by Senthil {\em et al.} \cite{flst1}. This phase can be realized for
$\TK\ll I$ in situations where either the inter-moment interactions $I$ are strongly
frustrated or where magnetic quantum fluctuations are strong. Then, the local moments may
form a paramagnetic spin liquid without broken symmetries, weakly interacting with the
$c$ electrons.
\footnote{
Paramagnetic states {\em with} broken symmetries, such as
valence-bond solids, generically lead to conventional phases, with a Fermi volume as in
the SDW and LMM phases. The same holds for situations with an even number of $f$
electrons in the crystallographic unit cell. }
Consequently, in this phase, dubbed fractionalized Fermi liquid FL$^\ast$,
the $f$ electrons do not contribute to the Fermi volume:
\begin{equation}
{\cal V}_{\rm FL^\ast} = K_d (n_c\,{\rm mod}\,2)
\end{equation}
Importantly, this ``small'' Fermi volume violates Luttinger's theorem,
hence this phase is a true non-Fermi liquid metal.
As discussed in Ref.~\cite{flst1}, the low-energy excitations of the
spin liquid, which are necessarily fractionalized, account for the Luttinger violation.
Thus, the Fermi volume provides a sharp distinction between FL and FL$^\ast$,
and a quantum phase transition must separate the two (Fig.~\ref{fig:doniach}b).
A more detailed discussion is given in the following sections.

In addition, other phases, e.g. with charge-density wave or superconducting order, can be
realized in the Kondo-lattice model, Eq.~(\ref{eq:KLM}), but are not of central interest in
this article.

\subsection{Kondo breakdown, fate of local moments, and global phase diagram}
\label{sec:global}

The experimental situation in a number of heavy-fermion metals
\cite{hvl,geg_rev,stewart01,stewart06} is that apparent quantum critical behavior is found near
the onset of a magnetic phase, with singularities being stronger than those predicted by
the LGW approach of Hertz. Initial speculations about a breakdown of the Kondo effect
\cite{schroeder98,schroeder00,coleman01} were followed by a number of concrete
calculations, to be reviewed below.
Conceptually, it is important to think about the relation between such a
Kondo-breakdown transition and long-range magnetic order. While early ideas
implied that the Kondo breakdown occurs concomittantly with the onset of long-range
antiferromagnetism -- in the terminology of the last section this is equivalent to a
direct transition from a paramagnetic heavy Fermi liquid (FL) to a local-moment magnetic
(LMM) metal -- this does not need to be the case.
To make this precise, it is useful to {\em assume} the existence of a Kondo-breakdown
transition and discuss both magnetism and the fate of the local moments in its vicinity.

If a system undergoes a Kondo-breakdown transition, the resulting non-Kondo (or OS Mott)
phase displays local moments which are not screened via the Kondo interaction with the
conduction band. (From the orbital-selective Mott perspective, local moments form in the
band undergoing Mott localization.) Then, the nature of the non-Kondo phase is determined
by how the spin entropy associated with these local moments is quenched in the
low-temperature limit. Conceptually, different possibilities can be envisioned, to be
discussed here for the most interesting case of one spin-1/2 moment per unit cell.
(ia) The local moments order either ferromagnetically or antiferromagnetically.
(ib) The system remains paramagnetic, but translation symmetry is broken such that the enlarged
unit cell contains an even number of spins 1/2 which combine to form a singlet
(equivalent to a valence-bond solid).
(ii) The local moments form a paramagnetic spin liquid without broken symmetries.
The discussion of the Fermi volumes in the preceding section shows that cases (ia) and
(ib) generically lead to conventional phases with Fermi-liquid-like properties, whereas
case (ii) necessarily results in a true metallic non-Fermi liquid phase.

The different non-Kondo phases naturally lead to different phase transition scenarios.
Cases (ia) and (ib) imply the existence of a symmetry-breaking phase transition, both at
finite temperature in the non-Kondo regime (if not prevented by the Mermin-Wagner theorem) and
at $T=0$. Imagine now a single non-thermal control parameter which allows to tune both
this symmetry-breaking transition and the Kondo-breakdown transition.
This allows us to distinguish two cases:
(A) Both zero-temperature transitions coincide (without further fine tuning).
(B) The transitions do not coincide. If the system enters the non-Kondo phase before the
onset of symmetry breaking, then the above case (ii) is realized in this intermediate
regime. If, on the other hand, the symmetry-breaking transition occurs first, the notion
of Kondo breakdown must be re-considered, because the Fermi volume count changes in the
presence of symmetry breaking, for a detailed discussion see Sec.~\ref{sec:af_kondo}.
If, as in case (ii), the non-Kondo phase does not break symmetries, there is only one
zero-temperature transition associated with Kondo breakdown. If a magnetic phase exists in some
enlarged space of control parameters, then we recover the transition scenario (B),
with separate transitions for Kondo breakdown and magnetism.

Scenario (A), with coinciding antiferromagnetic and Kondo-breakdown transitions, has been put
forward by Si and co-workers \cite{si01,si03}, with a concrete calculational scheme using a variant of
dynamical mean-field theory, see Sec.~\ref{sec:edmft}. Here it is argued that Kondo
screening is generically destroyed at an antiferromagnetic transition in two space dimensions,
i.e., the Kondo effect is driven critical by two-dimensional critical AF fluctuations.

In contrast, scenario (B) has been proposed first by Senthil and co-workers \cite{flst1},
in an attempt to disentangle the critical phenomena of Kondo breakdown and
antiferromagnetism. To this end, Ref.~\cite{flst1} assumed a situation where magnetic
order of the local moments is suppressed either by strong quantum fluctuations or by
geometric frustration. Then, a fractionalized Fermi liquid FL$^\ast$ is realized in the
regime of weak Kondo screening.
The transition between FL and FL$^\ast$ is not associated with a local order parameter or
symmetry breaking, but is accompanied by a jump in the Fermi volume (despite being a
continuous QPT).

\begin{figure}
\begin{center}
\includegraphics[%
  width=0.55\linewidth,
  keepaspectratio]{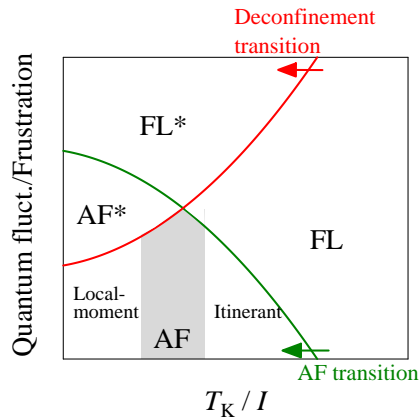}
\end{center}
\caption{
(Color online) ``Global'' zero-temperature phase diagram for a Kondo or
Anderson lattice (with one $f$ electron per crystallographic unit cell), showing two
transitions for the onset of antiferromagnetism and for the breakdown of the Kondo effect
(equivalently the onset of fractionalization).
FL$^\ast$ is the fractionalized Fermi-liquid phase, which is a metallic non-Fermi liquid
where the local moments form an exotic spin-liquid state.
Inside the AF phase, a crossover from more itinerant to more localized behavior occurs,
which may be accompanied by one or more transitions where the Fermi-surface topology
changes.
Finally, AF$^\ast$ is a fractionalized magnet, obtained from magnetic order in the spin
liquid of FL$^\ast$.
(Figure taken from Ref.~\cite{mv_af})
}
\label{fig:global}
\end{figure}

The key difference between the scenarios of Si {\em et al.} (A) and Senthil {\em
et al.} (B) is thus in the role of long-range ordered magnetism in destroying Kondo
screening: it is vital in the former, but unimportant in the latter case. In other words,
in scenario (B) Kondo screening is destroyed by competing {\it short-range} magnetic
fluctuations.
Consequently, this implies a difference in the resulting ``global''
zero-temperature phase diagram. In case (A), there are two different types of magnetic
transitions, depending on whether or not the onset of magnetism is accompanied by Kondo
breakdown, see Fig. 1 of Ref.~\cite{si_global}.
In contrast, in case (B) there are two distinct transition lines, one
associated with Kondo breakdown and fractionalization, the other one with magnetism. As a
result, there are four phases, Fig.~\ref{fig:global}, and the coincidence of magnetic and
Kondo-breakdown transitions requires fine tuning towards the multicritical point
\cite{flst1,mv_af}.
We note that, in principle, FL$^\ast$ may be unstable at lowest temperatures towards
magnetism (similarly, AF$^\ast$ may be unstable towards confinement) \cite{flst3}, as e.g. in the
scenario of deconfined criticality \cite{deconf}. However, the primary transition controlling the
critical behavior would still be FL--FL$^\ast$ in this case.
Recent experiments on doped \YRS\ \cite{friede09} have identified distinct
zero-temperature transition lines for antiferromagnetism and a Fermi-surface
reconstruction, which appears consistent with Fig.~\ref{fig:global}, for more details see
Sec.~\ref{sec:exp}.

So far, this discussion has focused on the limit of low temperatures where the entropy
associated with the magnetic moments is essentially quenched. However, the concept of OS
Mott transitions can make sense at elevated temperatures as well, namely if the Mott
scale is much larger than the scale of inter-site magnetic correlations. In other words,
if non-local exchange interactions are weak (e.g. due to some cancellation) there might
be an OS Mott crossover, separating a regime with well-defined weakly correlated local
moments, coexisting with metallic behavior, from a conventional Fermi-liquid regime.
Conceptually, this requires to replace non-local magnetic fluctuations as the driving
force of OS Mott physics by something else. As discussed in Sec.~\ref{sec:osmott}, this
role can be taken by intra-atomic Hund's rule correlations.

\subsection{Dynamical mean-field description of ``local'' criticality}
\label{sec:edmft}

An early approach designed to capture the breakdown of Kondo screening
due to magnetic bulk fluctuations employs an extension
of the dynamical mean-field theory (DMFT) and has been
worked out by Si {\em et al.} \cite{si01,si03}.
It led to the proposal of a ``local'' QCP (to be made precise below),
based on the idea that the breakdown of Kondo screening is a
spatially local phenomenon, i.e., it affects every spin of the
underlying Kondo lattice independently.

Within the extended DMFT (EDMFT), the Kondo-lattice model $\mathcal{H}_{\rm KLM} +
\mathcal{H}_I$, Eq.~(\ref{eq:KLM}),
is mapped to a so-called Bose-Fermi Kondo model with both a fermionic and
bosonic baths (represented by operators $c_k$ and $b_k$):
\begin{eqnarray}\label{HlEDMFT}
\mathcal{H}_{\rm loc}&=&\sum_{k \sigma} E_{k} c^\dagger_{k\sigma} c_{k\sigma}+
J \vec{S} \cdot \vec{s}_0
+
g \sum_k \vec{S} \cdot (\vec{b}_{k} +\vec{b}^\dagger_{-k})+
\sum_k \omega_k \vec{b}^\dagger_{k}\vec{b}_{k}.
\end{eqnarray}
Within EDMFT, the Green's functions and susceptibilities of the lattice model are
approximated by $1/g_{\vec{k}}(\w)\approx \w-\epsilon_{\vec{k}}-\Sigma(\w)$ and
$1/\chi_{\vec{q}}(\w)\approx I_{\vec{q}}+M(\w)$, where $\Sigma(\w)$ and $M(\w)$ are the
electron and boson self-energies of the local problem. The free parameters $E_k$, $\w_k$,
and $g$ are determined from the self-consistency condition that the local Green's
function and susceptibility in the global and local model, Eqs.~\ref{eq:KLM} and
\ref{HlEDMFT}, match. In other words, $c_k$ and $b_k$ represent the (local) fermionic and
magnetic degrees of freedom of the bulk.

The most important ingredient for local criticality is the behavior of the
Bose-Fermi Kondo model (\ref{HlEDMFT}). This impurity model displays a continuous
QPT between a phase with Kondo screening and one with universal local-moment
fluctuations \cite{mvrev}, provided that spectrum of the bosonic bath is gapless.
Within the EDMFT approach, the QCP of the lattice model (\ref{eq:KLM}) is thus
mapped onto the impurity QCP of Eq.~(\ref{HlEDMFT}). At the magnetic transition of the
lattice, the gapless bosonic bulk spectrum drives the Kondo effect critical.
At this ``local'' QCP all self-energies are momentum-independent,
and the non-local dynamics of the magnetic fluctuations is Gaussian.

A qualitative analysis reveals that this Kondo breakdown can occur only in $d=2$, whereas
in $d=3$ the local susceptibility is not sufficiently singular to render the Kondo effect
critical. In $d=2$, the logarithmic divergence of the local susceptibility at the QCP can
cause a power-law behavior of $M(\w)$ at $T=0$, $M(\w) = -I_{\vec{Q}} +
(-i\w/\Lambda)^\alpha$, where $\vec{Q}$ is the ordering vector, $\Lambda$ a cutoff, and
$\alpha$ a non-universal exponent \cite{si03}. A numerical solution of a simplified EDMFT
(without fermionic self-consistency and with Ising magnetic symmetry) has confirmed this
result \cite{isingedmft1,isingedmft2,isingedmft3}.\footnote{
Arguments have been put forward for a
generic first-order transition within EDMFT \cite{kotliar05}, the differences to
Refs.~\cite{isingedmft1,isingedmft2,isingedmft3} resulting from a different treatment of RKKY
interactions.
}
These results are in remarkable agreement with what has been found in the experiments
of \cite{schroeder00} on \CeAu, see Eq.~(\ref{chiAlm}) below -- in particular
the anomalous exponent of the susceptibility is obtained as $\alpha\approx 0.8$
while the value from fitting the experimental data is $\alpha\approx 0.74$.

To date, concrete predictions for quantum critical thermodynamic and transport properties
within this framework are lacking. Also, the role of magnetic frustration is unclear, as
there appears no room in this theory for a breakdown of Kondo screening without the onset of
magnetic order.

\subsection{Slave-particle theory of Kondo breakdown and fractionalized Fermi liquids}
\label{sec:sbmf}

A conceptually different approach to the breakdown of Kondo screening \cite{flst1} starts
by separating the phenomena of Kondo breakdown and ordered magnetism. When Kondo
screening breaks down {\em without} the simultaneous onset of magnetic order (or other
types of symmetry breaking), the resulting ground state is a paramagnet where the
conduction electrons form well-defined quasiparticles on their own and the local moments
are in a fractionalized (gapless or gapped) spin-liquid state -- this is the
fractionalized Fermi liquid (FL$^\ast$) advocated above.
These considerations lead to a modified Doniach phase diagram as in
Fig.~\ref{fig:doniach}b.

For concrete calculations, the transition from FL$^\ast$ to FL
can be analyzed in slave-boson mean-field theory plus
Gaussian fluctuations around the saddle point.
The mean-field Hamiltonian for the Kondo-lattice model $\mathcal{H}_{\rm KLM} +
\mathcal{H}_I$, Eq.~(\ref{eq:KLM}), reads \cite{flst1,flst2,burdin}
\begin{eqnarray}\label{mf1}
 \mathcal{H}_{\rm mf} & = & \sum_k \epsilon_k c^{\dagger}_{k\alpha} c_{k\alpha} -
  \chi_0\sum_{\langle rr' \rangle} \left(f^{\dagger}_{r\alpha} f_{r'\alpha} +
\mbox{
  h.c.}\right)
    \nonumber\\
  &
  + & \mu_f\sum_r f^{\dagger}_{r\alpha}f_{r\alpha}- b_0 \sum_k
\left(c^{\dagger}_{k\alpha}
  f_{k\alpha} + \mbox{h.c.} \right).
\end{eqnarray}
Here, $\vec S_r = \frac{1}{2} f^{\dagger}_{r\alpha} \vec \sigma_{\alpha\alpha'} f_{r\alpha'}$
is the auxiliary-fermion representation of the local moments.
Further, $\chi_0$, $\mu_f$, and $b_0$ are mean-field parameters, where $\chi_0$
represents inter-site correlations of local moments, and a non-zero $b_0$ signals Kondo
screening.
In the FL$^\ast$ phase, where $b_0=0$, the $f$ particles represent spinons of the
spin-liquid component.
A formally identical Hamiltonian emerges from a mean-field treatment of an Anderson
lattice model \cite{hewson,pepin07}.

On the mean-field level, the FL$^\ast$ to FL transition is signaled by the condensation of the slave
boson $b_0$ measuring the hybridization between the $c$ and $f$ bands. Close to the phase
transition, there are two Fermi surfaces of primarily $c$ and $f$ character, respectively.
Approaching the QPT from the FL side, the quasiparticle weight on an entire
sheet of the Fermi surface vanishes continuously. On the FL$^\ast$ side, this Fermi
surface becomes of Fermi surface of neutral $f$ spinons.
Importantly, this transition is {\em not} spatially local, as all self-energies retain
their momentum dependence.

The critical theory of the FL--FL$^\ast$ transition can be derived starting from the
slave-particle formulation, Eq.~(\ref{mf1}), supplemented by a gauge field. Provided that
the two Fermi surfaces do not overlap, the fermions can be integrated out, and one ends
up with a theory for dilute bosons $b$ coupled to a compact U(1) gauge field
\cite{flst2}. The transition is tuned by the chemical potential of the bosons; it occurs
at the (bosonic) wavevector $Q=0$ and has dynamical exponent $z=2$, it is thus above its
upper critical dimension. The FL phase corresponds to the confined phase (identical to
the Higgs phase) of the gauge theory, whereas the FL$^\ast$ phase represents the
deconfined phase \cite{compact}. Note that the coupling to the gauge field removes a
finite-temperature transition associated with deconfinement. The specific heat acquires a
singular contribution from gauge-field fluctuations with $C/T \sim \Log(1/T)$ in $d=3$ in
both the critical regime and in the FL$^\ast$ phase. Interestingly, the decay of the
bosons into particle--hole pairs becomes possible above an energy $E^\ast$ which can be
small if the momentum-space distance between the two Fermi surfaces is small; above this
energy the theory obeys $z=3$, and an additional $\Log(1/T)$ contribution in $C/T$
appears \cite{paul06}.
Following the initial proposal in Refs.~\cite{flst1,flst2}, a variety of physical
properties have been calculated from this critical theory, with results listed below in
Sec.~\ref{sec:obs}.

On lattices with underlying non-collinear magnetism, the gauge symmetry is reduced
from U(1) to Z$_2$. Then, the Z$_2$ spin-liquid component of the FL$^\ast$ phase displays
quite generically spinon pairing. It has been argued that this produces a robust
mechanism for superconductivity (which is almost certainly unconventional) masking the
FL--FL$^\ast$ transition \cite{flst1}.

For the mean-field theory in Eq.~(\ref{mf1}), the coupling to lattice degrees of freedom
has been considered in Ref.~\cite{hackl08}. The most interesting question is whether the
Kondo-breakdown transition can survive as a continuous transition upon coupling to
phonons, or whether it generically becomes discontinuous. As the critical hybridization
degrees of freedom couple linearly to strain, a first-order Kondo-breakdown could be
considered a quantum version of the celebrated Kondo volume collapse. The mean-field
analysis of Ref.~\cite{hackl08} shows that the FL--FL$^\ast$ transition remains
continuous for small electron--lattice coupling, but can become first-order for larger
coupling. In this case, it shows an interesting interplay with the Lifshitz transition
inside the FL phase.

Besides the mean-field theory in Eq.~(\ref{mf1}), based on slave bosons capturing Kondo
screening, a few other slave-particle theories have been proposed as starting point for a
description of Kondo-breakdown transitions, see e.g. Refs.~\cite{pepin05,kskim05,kskim10}.
However, most of these ideas have serious drawbacks, i.e.,
do not satisfactorily describe either the screened FL phase \cite{pepin05,kskim10} or the
Kondo-breakdown phase \cite{kskim05}. Hence, their predictive power is questionable.

\subsection{Kondo screening and Fermi surfaces in magnetically ordered phases}
\label{sec:af_kondo}

At the beginning of this section, we have argued that the Fermi volume provides a
clear-cut distinction between phases with and without Kondo screening. However, this is
only true in situations with spin-degenerate bands and an odd number of $f$ electrons per
unit cell, and hence does usually not apply to phases with magnetic long-range order.
Instead, the Fermi volumes of the SDW and LMM metallic magnets are identical. Therefore,
one has to ask whether such a sharp distinction remains in magnetically ordered phases,
in other words, whether the notion of the presence or absence of Kondo screening remains
precise. To be specific, the following discussion is given for antiferromagnet order
which enlarges the unit cell by a factor $N$, but applies similarly to valence-bond solid
order.

First, it is straightforward to see that the onset of magnetic order in a heavy FL will
lead to a reconstruction of the Fermi surface into pockets, provided that the ordering
wavevector connects pieces of the Fermi surface. Generically, the resulting SDW Fermi surface
has a different topology from the one of the bare conduction electrons, relevant to the
LMM metal.\footnote{
In the literature, the Fermi volumes of the SDW and LMM phases are sometimes dubbed
``large'' and ``small'', leading to phase labels \afl\ and \afs. However, the Fermi
surfaces differ only in topology, with the volumes being identical.
}
Then, the two phases must be separated by one or more Lifshitz transitions
where the Fermi-surface topology changes. However, counter-examples have been given in
Ref.~\cite{mv_af}, with identical Fermi-surface topologies in both limits and no need for
Lifshitz transitions.

Second, the question remains whether Kondo screening is sharply defined in the presence
of magnetic order. This issue is subtle as there is no gauge-invariant order parameter
for this problem. A recent renormalization-group treatment \cite{yama} staring from the
LMM phase in the decoupled limit found the Kondo coupling to be exactly marginal. While
the authors interpreted this in terms of the absence of Kondo screening \cite{yama}, a
different interpretation is suggested by the analogy to the Kondo effect in a magnetic
field. There, it is known that the Kondo coupling is marginal for any finite field, and
that a line of renormalization-group fixed points connects the screened impurity and the
fully polarized impurity without any phase transition. Hence, a sharp distinction between
the presence and absence of Kondo screening does not exist. Therefore, it is plausible
that the same applies to the antiferromagnetic Kondo lattice, i.e., Kondo screening
disappears smoothly in the antiferromagnetic phase \cite{mv_af}. Of course, this argument
does not exclude the existence of first-order transitions where both Fermi surface and
magnetism change discontinuously. Recent numerical results of microscopic Kondo-lattice
calculations \cite{assaad,fabrizio08}, which find that $c$ and $f$ bands are hybridized
on both sides of the topological transition involving a Fermi-surface reconstruction,
support this line of reasoning.

\subsection{Computational studies of Kondo and Anderson lattices}
\label{sec:cdmft}

The fate of Kondo screening across the heavy-fermion phase diagram can in principle be
assessed using computational techniques applied to Kondo or Anderson lattices. Here,
mainly variational techniques and cluster extensions of DMFT have been used. However, the
interesting case of geometric frustration is difficult to study due to the sign problem
in quantum Monte-Carlo approaches.

A $d=2$ square Kondo lattice was studied in Ref.~\cite{assaad} using cluster DMFT (with
cluster sizes up to 8 $f$ moments) as function of conduction band filling $n_c$ and the
ratio $J/t$ (where $t$ is the hopping matrix element). A single continuous transition was
found between a paramagnetic FL phase and an antiferromagnetic phase with a distinct
Fermi-surface topology (i.e. \afs), however, the numerical accuracy was not sufficient to
exclude the existence of two separate (i.e. magnetic and Lifshitz) transitions. Analyzing
the fermionic spectra in terms of mean-field band structures, the authors concluded that
Kondo screening is present also in the magnetic phase. Related results were obtained in a
variational Monte-Carlo study of the same model \cite{ogata}, but with a somewhat
different transition scenario. Depending on $n_c$, the authors found either a first-order
transition from FL to \afs, or a continuous transition from LM to \afl, followed by a
first-order transition to \afs. These results have been confirmed by a Gutzwiller
variational study \cite{fabrizio08}.

Two-site cluster DMFT has been used to further characterize the possible phase diagrams and
transitions \cite{deleo08a,deleo08b}. In particular, the OS Mott transition has been
studied in an Anderson model with magnetic order suppressed ``by hand'' \cite{deleo08b}.
Here, a continuous transition from a heavy FL to a paramagnetic non-Kondo phase is found,
in which the effective hybridization between $c$ and $f$ band vanishes in the low-energy
limit (but remains finite at elevated energies). This supports the notion that long-range
magnetism is not required for Kondo-breakdown (or OS Mott) physics.

It has to be kept in mind that cluster DMFT cannot reliably describe the critical
properties of the OS Mott transition, due to the finite cluster size. In fact, the
numerical results of Ref.~\cite{deleo08b} can be traced back to the behavior of the
two-impurity Kondo model which is known to display an interesting quantum phase
transition where Kondo screening disappears. However, this model has a residual entropy
at the QCP of $(\Log 2)/2$, which would imply an extensive entropy at the QCP in the
lattice model. As this cannot survive as $T\to 0$, the validity of the DMFT results is
restricted to elevated temperatures.


\section{Orbital-selective Mott transition in multi-band Hubbard models}
\label{sec:osmott}

The concept of an ``orbital-selective Mott transition'' was originally put forward in
Ref.~\cite{anisimov02} to explain the puzzling features of \CaSr\ (Sec.~\ref{sec:exp}).
This motivated a variety of computational studies of multi-band Hubbard models, mainly
using variants of DMFT, with yielded partially conflicting results due to differences in
input parameters. By now, a reasonably consistent picture has emerged, to be detailed
below.

In general, an orbital-selective Mott transition is a transition where one of
multiple correlated bands looses its metallic character, i.e., the electrons in one band
become Mott-localized in the ``orbital-selective Mott phase''. In a situation with
multiple Fermi sheets this implies that the quasiparticle weight on one of the sheets
vanishes at the transition. Apparently, the qualitative distinction between the
conventional metal and the OS Mott phase is again in the Fermi volume, which jumps by an
amount corresponding to a half-filled band upon crossing the transition (assuming a
single-site unit cell).

Similar to the considerations in Sec.~\ref{sec:global}, the fate of the OS Mott phase at low
temperatures is determined by the behavior of spin moments in the Mott-localized band.
If those moments order magnetically or produce a paramagnetic state with broken
lattice symmetry, then a conventional low-temperature phase results. If, in contrast,
a spin liquid without broken symmetries is realized, the OS Mott phase is a metallic
non-Fermi liquid of FL$^\ast$ type.

This discussion makes clear that there is no qualitative distinction between the
Kondo-breakdown QPT in heavy fermions and the OS Mott QPT in multi-band
Hubbard models. Rather, the Kondo-breakdown transition is a particular form of an OS Mott
transition where the electrons undergoing Mott localization are in the Kondo limit.
\footnote{
This observation was first formulated by C. P\'epin in Ref.~\cite{pepin07}.
The relation between Kondo effect, $f$ electron localization, and OS Mott physics was
discussed before by de Medici {\em et al.} \cite{medici04}.
}

\subsection{Computational studies of multi-band Hubbard models}

For multi-band Hubbard models, orbital-selective Mott transitions have been mainly
studied using DMFT techniques \cite{koga04,medici05,liebsch05,bluemer05,held05,millis07,costi07,medici09}.
In standard DMFT, the problem maps onto a self-consistent multi-level or multi-orbital impurity
model, where the OS Mott phase is characterized by a stable local moment in one of the
orbitals. Such a phase has been found in situations with a large ferromagnetic
inter-orbital (i.e. Hund's rule) coupling between inequivalent bands (i.e. with different
bandwidths and/or different correlation strengths).
Non-Fermi liquid properties of the OS
Mott phase at elevated temperatures, arising from the scattering of conduction electrons
on unscreened moments, have been studied in some detail \cite{medici05}.
The zero-temperature OS Mott transition is found to be either continuous or discontinuous, depending
on model parameters \cite{costi07}, and exists both for Heisenberg and Ising-symmetric
Hund's rule coupling \cite{bluemer05,costi07}.

\begin{figure}
\begin{center}
\includegraphics[%
  width=0.6\linewidth,
  keepaspectratio]{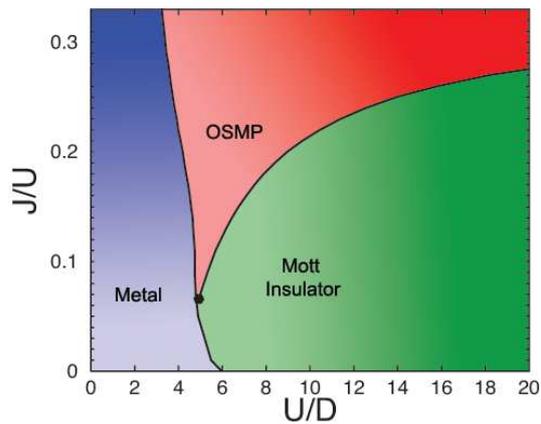}
\end{center}
\caption{
(Color online) Phase diagram of a three-band Hubbard model at total filling $n=4$, showing metallic,
Mott-insulating, and OS Mott phases. The bandwidths of all bands are identical, but one
band is lifted to higher energies by a crystal field splitting $\Delta$.
In the phase diagram, the band populations are fixed at (1, 1.5, 1.5) by adjusting
$\Delta$.
(Figure taken from Ref.~\cite{medici09})
}
\label{fig:luca}
\end{figure}

Recently, it has been emphasized that OS Mott phases can occur even in system with bands
of similar width and correlation strength, provided that the band degeneracy is lifted
(e.g. by crystal field splitting) and a sufficient Hund's rule coupling is present
\cite{medici09}, see Fig.~\ref{fig:luca}.
This open the possibility that OS Mott physics is a rather common
phenomenon in multi-orbital transition metal compounds.

As inter-site correlations are not captured in DMFT, a local-moment spin liquid (and
hence FL$^\ast$) cannot be described. Therefore, the low-temperature OS Mott phase of
DMFT is somewhat artificial (in the same sense as the DMFT Mott insulator is artificial):
there is a residual $\Log 2$ entropy per site, and the paramagnet state is unstable
towards magnetic order. One might in fact ask why the local moments present in the DMFT
OS Mott phase are not generically screened due to the presence of additional metallic
bands. The answer is that the relevant effective impurity model turns out to be a {\em
ferro}magnetic Kondo model, such that the two-fold degenerate state survives in the
$T\to0$ limit \cite{medici05}. Despite the absence of inter-site correlations, DMFT can
be expected to correctly capture local properties at elevated temperatures, such as the
evolution of the orbital occupations near an OS Mott transition. As discussed in
Sec.~\ref{sec:cdmft}, some of the DMFT deficiencies are cured in cluster versions of
DMFT; however, the correct quantum critical behavior cannot be recovered for finite
cluster size.

An important lesson from the DMFT studies is that Hund's rule
coupling tends to stabilize OS Mott phases. This constitutes a second driving force for
OS Mott transitions, i.e., Fermi-liquid-like screening of correlated electrons can be
suppressed both by {\em non-local} antiferromagnetic correlations and {\em local}
ferromagnetic correlations. While the first mechanism is required to stabilize the OS
Mott phase at low $T$ (i.e. to prevent magnetic order), the second one may be dominant in
multi-orbital situations at elevated temperatures (but is absent in standard Kondo
lattices).


\section{OS Mott phases and metallic spin liquids}
\label{sec:flst}

If the metallic phase resulting from an OS Mott transition out of a paramagnetic Fermi
liquid does not display magnetic long-range order or other types of spontaneous symmetry
breaking, it may be called a ``metallic spin liquid'', owing to the fact that the
electrons in the Mott-localized band now form localized spin moments. Microscopically,
the absence of magnetic order will be due to geometric frustration or enhanced quantum
fluctuations (e.g., by natural spin pairing due to structural dimers).

In the low-temperature limit, where the spin entropy is quenched, such a metallic spin
liquid can be characterized by its Fermi volume, which facilitates the distinction into
``conventional'' and ``unconventional'', according to whether Luttinger's theorem is
fulfilled or not.\footnote{
It may also be useful to use the term ``metallic spin
liquid'' for a crossover regime at finite temperatures, provided that charge fluctuations
are small. However, here the spin entropy may be a significant fraction of $\Log 2$.
}
For simplicity, we assume zero external magnetic field and spin-degenerate bands.

The ``conventional'' case is realized for an even number of $f$ electrons per unit cell.
As Luttinger's theorem is fulfilled, such a metallic spin liquid generically is a
conventional Fermi liquid. The Kondo-breakdown transition is a transition between two FL
phases, which have the same Fermi volume, but typically different Fermi surface
topologies. In the OS Mott phase sufficiently far away from criticality, the low-energy
quasiparticles will be light conduction electrons, and the local-moment sector will be
gapped. Hence, all low-temperature properties are that of a Fermi liquid (without
heavy-fermion like mass-enhancement). In the simplest case, the local moments form
singlets within each unit cell, but this is not required.
An explicit microscopic example is the bilayer Kondo model discussed in Ref.~\cite{tsmv},
where each crystallographic unit cell contains two $c$ and two $f$ electron orbitals.
In this case, it was explicitly shown that the Kondo-breakdown transition is masked by
a novel phase with inter-layer coherence, either in the particle--hole or in the
particle--particle channel. It is likely that this instability of the QCP is generic
feature.

In contrast, an odd number of $f$ electrons per unit cell leads to ``unconventional''
metallic spin liquids. Here, the notion of a ``small'' Fermi volume is clear-cut, i.e.,
Luttinger's theorem must be violated, see Sec.~\ref{sec:fvol}. Hence, a true non-Fermi
liquid phase emerges a low temperatures. According to our current understanding, the
paramagnetic spin liquid formed in the local-moment sector is inevitably topologically
ordered and displays fractionalized excitations of spinon type \cite{frust_sl}, which
co-exist with the Fermi-liquid-like excitations of the conduction band.\footnote{
The natural theoretical description involves, in addition to conduction electrons,
auxiliary fields for both spinons and the Kondo effect, coupled to a gauge field.
}
Consequently this non-Kondo phase was dubbed fractionalized Fermi liquid FL$^\ast$
\cite{flst1}.
We note that the proof of existence of such a phase simply follows from the proof of the
existence for a fractionalized spin liquid (see e.g.
Refs.~\cite{moessner01,triang_ring}): A Kondo lattice formed from conduction electrons
and such a spin liquid does necessarily realize an FL$^\ast$ phase for weak Kondo
coupling, because the spin liquid is a stable phase of matter and hence does not
qualitatively change its properties under weak perturbations.

Although our understanding of FL$^\ast$ phases is limited, a few properties can be
derived from general arguments. First, a direct measurement of the electronic Fermi
volume (e.g. by angle-resolved photoemission in a quasi-two-dimensional system) is
possible in principle and would directly show the violation of Luttinger's theorem.
A somewhat less stringent criterion is the {\em absence} of a certain band in
photoemission at low energies, which is predicted to cross the Fermi surface from ab-initio
calculations.

Second, specific thermodynamic and transport properties characterize FL$^\ast$ phases.
If the local-moment sector is fully gapped in both the
singlet and triplet channels, as is the case for Z$_2$ fractionalization, then most
low-temperature properties are Fermi-liquid-like. However, the topological order implies
non-trivial response to boundaries and impurities as well as unusual finite-temperature
crossovers. In contrast, much stronger violations of Fermi-liquid behavior occur for
gapless spin liquids with U(1) fractionalization. Here, low-energy excitations may exist
in both the spinon and gauge-field sectors.
A few concrete results are available for the case of a U(1) FL$^\ast$ phase with a spinon
Fermi surface \cite{flst2}:
the specific heat obeys $C \propto T \Log (1/T)$ in $d=3$ and $C \propto T^{2/3}$ in
$d=2$ from gauge fluctuations.
Electrical transport is dominated by $c$ electrons and hence of Fermi-liquid form, i.e.,
the resistivity is given by $\rho = \rho_0 + A T^2$.
For thermal transport, a careful distinction is required between the impurity-dominated
low-$T$ regime (``dirty limit'') and the ``clean limit'' at elevated $T$.
The spinon thermal conductivity in the clean limit is given by $\kappa_s \propto T^{(3-d)/3}$
whereas it is $\kappa_s \propto T$ in the dirty limit \cite{nave,hackl10}.
The total thermal conductivity is a sum of spinon and electron contributions, where the
latter has standard Fermi-liquid form. This implies that the Wiedemann-Franz law,
\begin{equation}
\label{wflaw}
\frac{\kappa}{\sigma T} = L_0, ~~L_0 = \frac{\pi^2}{3} \frac{k_B^2}{e^2},
\end{equation}
is violated in the $T\to 0$ limit, as both spin-liquid and conduction-electron components
contribute to heat transport. In addition, the low-$T$ corrections to $\kappa/T = {\rm
const}$ will be non-Fermi-liquid like -- recall that electrical transport is
Fermi-liquid like at low $T$.

As already stated at the end of Sec.~\ref{sec:global}, the concept of OS Mott phases may
also be well defined at elevated temperatures where the spin entropy is not fully
quenched, e.g., in a multi-orbital Hubbard system with large Hund's rule coupling and
small inter-site exchange. There, the typical signature of OS Mott physics would be the
co-existence of metallic resistivity and Curie-like local-moment susceptibility. However,
this notion is not clear-cut: a Kondo-lattice system in its high-temperature incoherent
regime above the Fermi-liquid coherence temperature may display similar characteristics,
but would not deserve the term ``OS Mott''.


\section{Critical Fermi surfaces}
\label{sec:FS}

A particularly fascinating aspect of orbital-selective Mott transitions is that of
``critical'' Fermi surfaces at the transition point. As described above, crossing the
transition at $T=0$ is accompanied by a jump in the Fermi volume. This is compatible with
a continuous QPT if the quasiparticle weight $Z$ at (a sheet of) the Fermi surface
vanishes continuously upon approaching the QCP. At the QCP, $Z=0$, and the pole in the
single-particle Green's function can be expected to be replaced by a power-law
singularity. Hence, the location of the Fermi sheet is well-defined in momentum space
through this singularity (despite $Z=0$) -- this has some similarity to Luttinger liquids
in $d=1$ spatial dimensions.

In the slave-particle theory described in Sec.~\ref{sec:sbmf} there are two Fermi sheets
in the FL phase, and, upon approaching the QCP, $Z$ goes to zero on one of those sheets,
while the other sheet is non-singular across the transition and takes the role of the ``small''
Fermi surface of the FL$^\ast$ phase. Hence, at the QCP the Fermi volume is still ``large''
in the sense of ${\cal V}_{\rm QCP} = K_d [(n_c + n_f)\,{\rm mod}\,2]$.
Alternatively, one could envision a scenario with a single ``large'' Fermi sheet in the FL
phase and a different ``small'' Fermi sheet in the FL$^\ast$ phase, both becoming singular upon
approaching the QCP from either side. Then, the QCP would be characterized by a
``super-large'' Fermi surface with ${\cal V}_{\rm QCP} = K_d [(2n_c + n_f)\,{\rm
mod}\,2]$ \cite{senthil06}. A mean-field theory for this interesting possibility is not known.
We also note that, at a continuous Mott metal--insulator transition, one can similarly expect
a vanishing quasiparticle weight on the entire Fermi surface.

In all cases, the state at the QCP is a critical non-Fermi liquid where the defining
characteristic of a metal, namely the Fermi surface of quasiparticles, is destroyed. A
phenomenological scaling framework for such states has been worked out by Senthil
\cite{senthil08}.
Concretely, the zero-temperature single-particle spectrum $A(\vec k,\w)$
can be expected to follow the scaling form \cite{senthil08}
\begin{equation}
A(\vec k,\w) \propto \frac{1}{\w^{\alpha/z}} F\left(\frac{\w}{(c k_{\perp})^z}\right)
\end{equation}
where $k_\perp$ measures the distance to the Fermi surface, $\alpha$ and $z$ are universal
exponents, and $c$ is a non-universal velocity.  In particular, it is permissible to have
exponents $\alpha$ and $z$ which are direction-dependent, i.e., vary along the Fermi surface.
This allows for a wide variety of critical thermodynamic behavior. Unfortunately, our
microscopic understanding of critical Fermi surfaces is rather limited: to our knowledge,
the only explicit results are those obtained within the framework of slave-particle plus
gauge-field theories as in Ref.~\cite{senthil08b}.
Clearly, more theoretical work is required in this direction, and should also be
corroborated by high-resolution ARPES experiments on candidate materials for continuous
Mott or OS Mott transitions.

Furthermore, there is the exciting possibility that a strongly singular Fermi surface
might render the corresponding QCP unstable towards a novel phase where the Fermi surface
is removed, e.g., by superconducting pairing. This constitutes a novel mechanism for
superconductivity driven by quantum criticality, but to our knowledge explicit
calculations illustrating this scenario are not available.


\section{Results for observables near quantum criticality}
\label{sec:obs}

In this section, we list some results for thermodynamic and transport properties near
orbital-selective Mott/Kondo-breakdown phase transitions which have been derived from the
available theoretical descriptions. We also discuss which experimental signatures may
allow to distinguish OS Mott transition from other candidates.

\subsection{Quantum critical regime}

To date, two conceptually distinct descriptions of OS Mott/Kondo-breakdown quantum
criticality are available, namely the EDMFT formulation of Sec.~\ref{sec:edmft} and the
slave-particle theory of Sec.~\ref{sec:sbmf}. Here, we shall reproduce important results
which have been obtained along these two lines. Before, a few qualitative statements are
in order. First, in heavy fermions the breakdown of Kondo screening does {\em not} imply
that the local moments are free to fluctuate at the QCP; instead anomalous power laws in
spin correlations will appear. Similarly, there will be no $\Log 2$ entropy per spin at
the QCP or in the quantum critical region. Second, as usual in quantum critical
phenomenology, all observables will display an extended crossover regime at finite $T$
from their large-Fermi-surface properties to the ones associated with the small Fermi
surface, i.e., two crossover lines will bound a quantum critical regime associated with
critical (or fluctuating) Fermi surfaces, Fig.~\ref{fig:doniach}b.

The predictions of the EDMFT approach, where Kondo-breakdown and magnetic transitions coincide,
are mainly restricted to the behavior of the magnetic susceptibility. At criticality in $d=2$
(recall that there is no Kondo breakdown in $d=3$ within EMDFT),
it is of the form $1/\chi(\vec{q},\w) = f(\vec{q})+(-i \w/\Lambda)^{\alpha}$ where
$f(\vec{q})$ vanishes at the ordering wavevector $\vec Q$, and the anomalous exponent was
found as $\alpha\approx 0.8$ \cite{isingedmft1,isingedmft2,isingedmft3}.
In this form of $\chi$, momentum and frequency dependence
``separate'', very similar to the experimental results on \CeAu, see Eq.~(\ref{chiAlm})
in Sec.~\ref{sec:exp} below.

Within the slave-particle theory for the FL--FL$^\ast$ transition, with fermionic
spinons and a U(1) gauge field \cite{flst2}, various thermodynamic
and transport properties have been calculated. We note that a similar effective theory
emerges for a filling-controlled Mott transition \cite{senthil08b}, with the main
difference that the $c$ band is absent here.
For the two-band heavy-fermion case, it has been pointed out that
a crossover energy scale $E^\ast$ emerges in the quantum critical regime, where the
critical dynamics changes \cite{paul06,paul08}. Below $E^\ast$, the Kondo bosons are
undamped and the theory has dynamical exponent $z=2$, whereas above $E^\ast$ the bosons
can decay into pairs of $c$ and $f$ particles, such that $z=3$. The value of $E^\ast$ is
set by the momentum-space distance of the $c$ and $f$ Fermi surfaces, and has been argued
to be generically small in heavy fermions \cite{paul06}. All the following results are
given for $d=3$ unless otherwise noted.

The quantum critical crossover lines, Fig.~\ref{fig:doniach}b, obey $T^\ast\propto|r|^{2/3}$ in the
asymptotic low-temperature regime, $T<E^\ast$, as the boson self-energy is
$\Sigma_b(0,0)\propto T^{3/2}$ \cite{flst2}. For $T>E^\ast$, this is replaced by
$T^\ast\propto|r|^{3/4}$ \cite{paul08}. Notably, there is a second crossover line on the
metallic side at a smaller $T^{\ast\ast} \propto {T^\ast}^{3/2}$, and Fermi-liquid
behavior is only established for $T<T^{\ast\ast}$. Physically, the condensation of Kondo
bosons happens below the scale $T^\ast$, but the spinon-gauge system remains critical
above $T^{\ast\ast}$. The regime $T^{\ast\ast} < T < T^\ast$ thus displays
non-Fermi-liquid behavior as well, for details see Ref.~\cite{senthil08b}.

We now turn to properties in the quantum critical regime, $T>T^\ast$. At
low temperatures, $T<E^\ast$, the specific heat follows $C \propto T \Log(1/T)$
from transverse gauge fluctuations \cite{flst2}. The same applies to $T>E^\ast$, where an
additional contribution of similar form comes from the Kondo bosons. Weakly singular
corrections occur to the static uniform susceptibility: $\delta\chi\propto -T^2\Log(1/T)$
for $T<E^\ast$ and $\delta\chi\propto -T^{4/3}$ for $T>E^\ast$ \cite{paul08}. The
Gr\"uneisen ratio of thermal expansion and specific heat, $\Gamma=\alpha/C_p$,
which is generically divergent at a QCP \cite{markus}, has been found to follow $\Gamma
\propto T^{-2/3} / \Log T$ for $T>E^\ast$ \cite{pepin08}.

Calculations of quantum critical transport are notoriously difficult, due to the
interplay of inelastic critical and elastic impurity scattering. This can induce a
variety of non-trivial crossovers already for LGW-type theories, as has been shown for
the metallic antiferromagnet by Rosch \cite{rosch00}. Moreover, most quantum critical
transport calculations {\em assume} that the subsystem of critical bosons is in
equilibrium, thus providing a mechanism for current relaxation. Such an assumption
implies that impurity scattering is rather effective in equilibrating the bosons, but this piece
of physics is not explicitly treated.
For the Kondo-breakdown field theory, the electrical conductivity may be derived from the
conductivities of the subsystems, together with Ioffe-Larkin composition rules
\cite{flst2,paul08,ioffe}. For $T<E^\ast$, the boson and spinon conductivities are
$\sigma_b \propto \Log (1/T)$ and $\sigma_f \propto T^{-5/3}$, respectively, and both
divergencies will be cutoff by residual impurity scattering. For the final result, one
again has to distinguish the low-$T$ ``dirty'' limit, where the temperature dependence
leads to small corrections to a constant dc conductivity, and the ``clean'' limit where
impurities can be ignored. In the dirty limit, the temperature dependence is
dominated by the bosons, $\delta\rho \propto \Log(1/T)$ \cite{flst2}. However, depending
on prefactors, this weak logarithmic dependence may easily be unobservable. In the clean
limit, the conduction electrons dominate, leading to the usual Fermi-liquid resistivity.
At elevated temperatures, $T>E^\ast$, the conduction electrons acquire an additional
scattering channel and generically dominate the $T$ dependence, with $\delta\rho \propto
T\Log(1/T)$ \cite{paul06,paul08}. Thermal transport has been considered in the same
framework, and $(\kappa/T)^{-1}$ was found to follow a similar $T\Log(1/T)$ dependence,
but here both $c$ and $f$ particles contribute. As a result, the Wiedemann-Franz law
(\ref{wflaw}) is violated both in the limit $T\to 0$ and at elevated $T$ \cite{pepin09}.

In addition, low-temperature transport properties {\em outside} the quantum critical
regime have been argued to jump when crossing the QCP from FL to FL$^\ast$
\cite{coleman01,si03,coleman05}, which is plausible considering the reconstruction of the
entire Fermi surface. An explicit calculation of longitudinal and Hall conductivities is
in Ref.~\cite{coleman05}. The low-$T$ optical conductivity near the FL--FL$^\ast$
transition has been calculated in Ref.~\cite{paul10} at the mean-field level. The result
shows that the typical mid-infrared feature, resulting from inter-band transitions in the
heavy FL phase, moves in a characteristic fashion to low energies upon approaching the
transition from the FL side, before its intensity diminishes.

A number of the listed results for $T>E^\ast$ are in agreement with experimental data on \YRS, at least
in certain regimes of temperature -- this supports the notion that Kondo breakdown
physics is realized in this material. However, there are a few caveats in this
interpretation, see Sec.~\ref{sec:exp}.


Finally, it should be noted that, by construction, antiferromagnetic fluctuations are not
critical near the FL--FL$^\ast$ transition. The interplay of this QCP with
additional magnetic quantum criticality has not been thoroughly investigated, but could
be highly relevant experimentally.

\subsection{Qualitative distinctions: Kondo breakdown vs. other phase transitions}

Unusual power laws in observables as listed above, restricted to a regime which narrows
in control parameter space as temperature is decreased, are certainly suggestive of
a continuous orbital-selective Mott transition. However, care is required for several reasons:
(a) Power laws extracted from experimental data typically cover one to two decades (e.g.
in temperature) only. Moreover, there is often a zero-temperature offset (as in
resistivity) or a background (as in susceptibility from nuclear moments)
which needs to be subtracted. As a result, it may be difficult to distinguish asymptotic
power laws from crossover phenomena.
(b) For certain observables, other quantum phase transitions may display similar
characteristics. For instance, a logarithmically diverging specific-heat coefficient is
expected for both itinerant antiferromagnetic transitions in $d=2$ and for OS Mott transitions
involving a U(1) gauge field in $d=3$.
(c) The interplay of quenched disorder and criticality possibly also induces unconventional power
laws, in principle even with exponents varying as function of a control parameter.

Therefore, it is worth discussing what could be considered as smoking-gun evidence for OS
Mott transitions. The most important aspect of an OS Mott transition is the collapse of
the Fermi surface -- in fact, this may be used as a {\em defining} criterion for such a
transition.
It implies a Fermi-surface reconstruction {\em across} the transition,
which is detectable via low-temperature measurements: a jump in the longitudinal and
Hall resistivities is expected \cite{coleman01,si03,coleman05},\footnote{
At a SDW transition the Fermi surface evolves
continuously, and the Hall coefficient displays a kink, but no jump;
only at a magnetic-field driven transition does the derivative of the
Hall current with respect to the magnetic field jump.
}
as well as an abrupt change in
de-Haas-van-Alphen or Shubnikov-de-Haas oscillations upon crossing the phase transition
in the limit $T\to 0$.
\footnote{
Angle-resolved photoemission (ARPES) experiments can in principle detect the Fermi-surface reconstruction
as well, but the present-day energy resolution is often insufficient for this purpose, in
particular in the context of heavy-fermion materials.
}
Of course, care is required here as well: On the one hand, abrupt changes may simply
arise from a first-order transition, e.g., between a paramagnetic and an
antiferromagnetic Fermi liquid. Hence, unconventional power laws are still
required to make the case. On the other hand, the extrapolation of finite-temperature
data to $T=0$ may be non-trivial, in particular in the presence of additional low-lying
energy scales in the system under consideration.

Measurements {\em inside} the putative OS Mott phase can only give clear-cut evidence in
the absence of symmetry breaking (in particular, of magnetism), as in the {\em presence}
of magnetic order the sharp distinction between small and large Fermi surfaces
disappears, as does the sharp distinction between the presence and absence of Kondo
screening (Sec.~\ref{sec:af_kondo}). Therefore, off criticality, it requires detecting a
paramagnetic non-Fermi liquid phase (Sec.~\ref{sec:flst}) to obtain evidence for
an OS Mott transition (which, however, could still be of first order).

In summary, both the abrupt Fermi-surface reconstruction across the transition and the
existence of a paramagnetic non-Fermi liquid phase at lowest $T$ are solid indicators
of OS Mott phenomena.


\section{Candidate materials for orbital-selective Mott transitions}
\label{sec:exp}

In this section, we give a brief overview on possible experimental realizations of
orbital-selective Mott physics, i.e., materials where experimental data have been
interpreted in these terms. The list includes both 4f and 5f rare-earth compounds and transition
metal oxides.

\subsection{\CeAu}

\CeAu\ was the first heavy-fermion system with a tunable quantum critical point, where
deviations from LGW behavior could be convincingly demonstrated \cite{hvl}. At the QCP from a
paramagnetic to an antiferromagnet metal, located at at $x_c=0.1$, the specific heat
coefficient $\gamma = C/T$ diverges logarithmically between 0.05 and 2.5\,K,
$\gamma = a \Log(T_0/T)$ \cite{hvl96,hvl98},
and the resistivity follows $\delta \rho \propto T$ \cite{hvl98b}.
So far, these properties are consistent with 2d antiferromagnetic LGW criticality,
and, remarkably, 2d spin critical fluctuations have indeed been detected in neutron
scattering \cite{stockert98}.
On top of this, the dynamic structure factor at criticality has been found to obey the
scaling law \cite{schroeder98,schroeder00}
\begin{equation}\label{chiAlm}
\chi^{-1}(\vec q,E,T) = c^{-1} (f(\vec q) + (- iE + aT)^{\alpha})
\end{equation}
with $\alpha\approx 0.8$.
This scaling is clearly not consistent with LGW criticality, because
(i) $\w/T$ scaling can only be expected for $d<2$ and
(ii) Eq.~(\ref{chiAlm}) implies anomalous behavior at {\it all} wavevectors, including
$\vec q = 0$ -- the latter has indeed been verified in measurements of the uniform
susceptibility.
In fact, it was the scaling law (\ref{chiAlm}) which fueled ideas of ``local''
criticality, see Sec.~\ref{sec:edmft}.

To date, \CeAu\ remains one of the best characterized heavy-fermion systems exhibiting
NFL behavior. Whether it indeed represents a realization of a Kondo-breakdown transition
is not known beyond doubt. While the behavior of $\gamma$, $\rho$, and $\chi$ at $x_c=0.1$
might be consistent with this hypothesis, there are no clear indications of a vanishing
thermodynamic energy scale at the QCP, and jump-like features in low-temperature
transport properties across the QCP have not been reported.
In fact, several transport experiments inside the antiferromagnetic phase \cite{hvl98,wilhelm}
are nicely explained by a competition between magnetic order and Kondo screening,
which may be taken as evidence {\em against} a Kondo breakdown scenario.
A further (perhaps unrelated) puzzle concerns the
large specific heat coefficient in the ordered phase: e.g. at 100 mK $\gamma$ is larger
for a range of concentrations above $x_c$ as compared to $x_c$ \cite{hvl96,hvl98}.
Here, it might be necessary to consider quenched disorder as an additional player.

\subsection{\YRS}

\YRS\ is the second prominent heavy-fermion metal where signatures of unconventional quantum
criticality have been found \cite{trovarelli00,gegenwart02}.
The stoichiometric compound displays an ordering transition at 70 mK to a
phase which is believed to be antiferromagnetic. This transition can be suppressed by
application of a small in-plane field of $\mu_0 H_N \approx 60$\,mT.
For small fields $H\lesssim H_N$, the specific heat coefficient varies logarithmically
between 0.3 and 10\,K,
while a stronger divergence is found below 0.3\,K (which is cut-off by magnetic
order for $H<H_N$) \cite{gegenwart02}. The resistivity varies linearly with temperature
up to 1\,K. These apparent similarities to \CeAu\ suggest that their quantum phase
transitions belong to the same universality class.
However, no neutron scattering data are available for \YRS\ to date.

Interestingly, a set of subsequent experiments supported the idea of a Fermi-surface
reconstruction near the QCP. Magneto-transport measurements showed a distinct crossover
in the Hall coefficient upon variation of the magnetic tuning field across the QCP
\cite{paschen04,friede10}, i.e., along a line $T_{\rm Hall}(H)$ which approximately
terminates at $H_N$. The magnetic-field width of the crossover is reported to vary linearly with temperature
\cite{friede10}, such that the extrapolation to $T=0$ results in a jump in the Hall
coefficient as function of $H$.
Further, maxima in the ac susceptibility and kinks in magneto-striction data \cite{gegenwart07}
similarly allowed to trace a line $T^\ast(H)$ which essentially coincides with $T_{\rm Hall}(H)$
and indicates the vanishing of one or several energy scales at the QCP.
These features appear consistent with the scenario of a Kondo-breakdown transition,
although a detailed understanding of the behavior near $T^\ast(H)$ is lacking.

\begin{figure}
\begin{center}
\includegraphics[%
  width=0.6\linewidth,
  keepaspectratio]{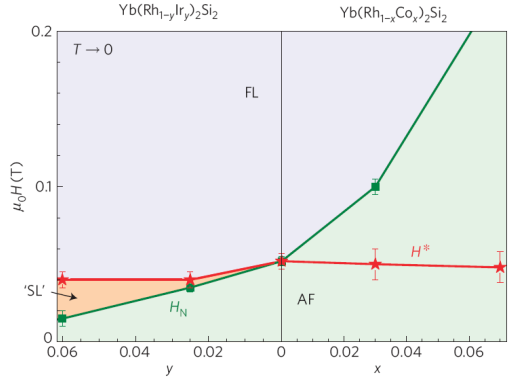}
\end{center}
\caption{
(Color online) Experimental phase diagram of doped \YRS, extrapolated to zero temperature.
The red line represents the critical field of $T^\ast(H)$
and the green line the antiferromagnetic (AF) critical field $H_N$. The blue and
green regions mark the Fermi-liquid (FL) and magnetically ordered ground
state, respectively. The orange region situated between these two is a
phase that may represent a metallic spin liquid (SL).
(Figure taken from Ref.~\cite{friede09})
}
\label{fig:friede}
\end{figure}

Exciting results have been very recently obtained from investigations of doped \YRS. In
both Yb(Rh,Ir)$_2$Si$_2$ \cite{friede09} and YbRh$_2$(Si,Ge)$_2$ \cite{custers10},
magnetic order is suppressed (i.e. the critical magnetic field $H_N$ decreases) compared to
stoichiometric \YRS, while it is enhanced in Yb(Rh,Co)$_2$Si$_2$ \cite{friede09} -- this
can be understood as an effect of chemical pressure. Remarkably, the $T^\ast(H)$ line,
which could be identified in all doped compounds, appears rather insensitive to doping,
such that its $T=0$ endpoint is now located at a field different from where magnetic
order disappears. Thus, the apparent Fermi-surface reconstruction and the onset of
magnetic order can be separated by doping (or pressure), see Fig.~\ref{fig:friede}.
If the Kondo-breakdown interpretation is correct, then this lends strong support to the
global phase diagram in Fig.~\ref{fig:global}, and suggests that both Ge- and Ir-doped
\YRS\ show a metallic non-Fermi liquid phase at small fields. In fact, recent transport
data on Ge-doped \YRS\ appear consistent with this assertion \cite{custers10}.

It must, however, be mentioned that a few properties of \YRS\ are not easily consistent
with the Kondo-breakdown scenario. First, it is puzzling that $T^\ast(H)$ is little
sensitive to doping: if it results from a competition of Kondo screening and magnetic
order (the latter being suppressed by the field), then it should display a significant
dependence on doping or pressure, both of which influencing hybridization matrix
elements. Second, it is equally puzzling that signatures of $T^\ast$ never appear in zero
field. Whether the tendency of the material towards ferromagnetism is relevant here is
not known. Third, the linear-in-$T$ width of the Hall crossover \cite{friede10} suggests
that the critical exponents obey $\nu z =1$, inconsistent with the results from
Gr\"uneisen measurements which point towards $\nu z = 0.7$ \cite{kuechler03}. (However, a
full transport calculation across the phase diagram, which would provide a prediction for
the crossover width, is lacking.) Finally, the role of the 0.3 K scale present in the
specific heat data is unclear.

\subsection{CeRhIn$_5$}

The antiferromagnet metal CeRhIn$_5$ belongs to the Ce-115 family of heavy fermions. It
can be driven into a superconducting phase by hydrostatic pressure, with a region of
coexistence in between \cite{park06,park08}. Upon additionally applying a magnetic field, there
is evidence for a single QCP between antiferromagnetic to non-magnetic phases
\cite{park06,knebel08}. Indeed, de-Haas-van-Alphen measurements in fields up to 17\,T
detected a change in the Fermi surface properties at a critical pressure of $p_c \approx
2.3$ GPa \cite{shishido}. Thus, the experimental data may be consistent with a
Kondo-breakdown transition upon lowering $p$, which occurs concomitantly with the onset
of AF order, but additional experimental data in this regime is needed.

Evidence for a transition inside an AF phase, with a Fermi surface reconstruction, has
been recently found \cite{goh} in CeRh$_{1-x}$Co$_{x}$In$_{5}$. This transition, however,
is strongly first order and also accompanied by a change in the magnetic structure.

\subsection{Uranium heavy fermions}

Uranium's 5f electron are known to be less strongly correlated than the 4f electrons of
rare earths like Cerium or Ytterbium. This has lead to suggestions about a ``dual''
nature of 5f electrons in compounds like UPt$_3$ and UPd$_2$Al$_3$
\cite{sato01,zwick02,zwick03}. Under the phenomenological assumption that one itinerant and
two localized $f$ electrons co-exist, important features like heavy quasiparticles with a
distinct mass anisotropy could be explained. Cluster calculations taking into account
intra-atomic correlations have provided a partial justification for this assumption
\cite{zwick04}.

Naturally, these considerations can be cast into the framework of OS Mott phases and
transitions. Indeed, recent theoretical work \cite{burdin10} studied a microscopic model relevant for
UPt$_3$ and constructed a mean-field phase diagram, which displays (depending on
correlation strength and bandwidths) a variety of partially localized phases. A
microscopic experimental verification would certainly be exciting.

\subsection{\CaSr}

The substitution of Ca for Sr in the unconventional superconductor Sr$_2$RuO$_4$ drives
the material towards a more correlated state, with Ca$_2$RuO$_4$ being an antiferromagnetic
Mott insulator with $S=1$ moments on Ru \cite{naka00}. The electronic
properties of \CaSr\ evolve non-monotonically with doping, with insulating behavior
appearing for $x<0.2$. Highly interesting is the physics at $x=0.5$ where the
susceptibility indicates almost free spin-1/2 moments while transport properties are
metallic.

This has prompted Anisimov {\em et al.} \cite{anisimov02} to propose an orbital-selective
Mott scenario for \CaSr, where localized and itinerant Ru 4d electrons co-exist for
$0.2<x\leq 0.5$. They supported this idea by ab-initio calculations which show that Ca
substitution leads to electron transfer from the wide $d_{xy}$ to the narrow $d_{xz,yz}$
bands, which (together with a rotation of RuO$_6$ octahedra) can cause partial Mott
localization in the two degenerate $d_{xz,yz}$ bands. For the OS Mott regime, it was
argued that ferromagnetic exchange competes with antiferromagnetic RKKY interactions,
suppressing magnetic order down to very low temperatures.
We note, however, that different scenarios for the doping evolution of the electronic
structure of \CaSr\ have been put forward \cite{liebsch07}, and more experiments are
required to investigate the intriguing properties of the $x<0.5$ region.

Notably, recent photoemission data \cite{neupane09} obtained on $x=0.2$ samples of \CaSr\
point towards a more complicated OS Mott scenario: There, the wider $d_{xy}$ band was found to
be absent from the Fermi surface. This was attributed to partial Mott localization of
this band in the presence of a $\sqrt{2} \times \sqrt{2}$ reconstruction due to rotation
of the RuO$_6$ octahedra which doubles the unit cell. The emerging six-band situation has
not been investigated theoretically, but the authors have supported their ideas using
mean-field treatments of simplified three-band models \cite{neupane09}.

\subsection{He bilayers}

Over the past decades, beautiful experiments on layers of fermionic He$^3$ adsorbed on
graphite have uncovered interesting correlation effects
\cite{greywall90,saunders97,saunders07}. For a single layer of He$^3$, solidification
occurs at 4/7 coverage \cite{greywall90}, which has been associated to a fermionic Mott
transition \cite{casey03}. The spin degrees of freedom are believed to form a spin liquid
dominated by ring-exchange interactions \cite{triang_ring,saunders97,roger90}.

Those experiments have been extended to He$^3$ bilayers, where the top layer can exist in
contact with a bottom layer which is not yet solidified. As function of the total
coverage, an interesting phase diagram has been experimentally determined, which displays
two apparent phase transitions \cite{saunders07}. It has been proposed that this physics
can be understood in terms of an Anderson lattice model, where the weakly correlated top
layer takes the role of $c$ electrons, while the strongly correlated bottom layer
contains $f$ electrons \cite{saunders07,benlagra08}. Naturally, this system is an ideal
candidate for an orbital-selective Mott transition, as the correlated bottom layer can
realize a paramagnetic spin liquid. A detailed modelling has been proposed in
Ref.~\cite{benlagra08}, employing a fermionic mean-field theory similar to that of
Ref.~\cite{flst2} and adapted to the Anderson model. This theory, assuming the existence
of an orbital-selective Mott transition, successfully explains some but not all key
features of the experiment \cite{benlagra08,benlagra10}.
It also displays a rather sharp change of the band structure
near criticality, which likely renders the quantum critical regime of the OS Mott
transition hard to observe.

\subsection{Cuprate superconductors}

Cuprate high-temperature superconductors constitute one of the most fascinating
topics in condensed matter physics. Despite almost 25 years of research, both the
mechanism underlying the superconductivity and the unusual normal-state properties are
still not fully understood. Conceptually, these difficulties arise from strong local
correlations which drive the parent compounds to be Mott insulators; understanding the
behavior of Mott insulators upon carrier doping poses a serious challenge to theory \cite{htsc_rev}.

Computer simulations of the Hubbard model \cite{haule07,jarrell08b,imada08,werner09,ferrero09}
using cluster extensions of DMFT have provided some insight into phenomena which are
of relevance for doped Mott insulators. Among other things, these calculations
show pseudogap behavior at low doping in the absence of superconductivity or magnetic
order. In the single-particle Green's function, this pseudogap behavior is accompanied by
the disappearance of the Fermi surface in parts of the Brillouin zone, namely in the
so-called antinodal regions near $(\pi,0)$, $(0,\pi)$ \cite{imada08,werner09,ferrero09}.
This low-doping
behavior is found to be qualitatively distinct from that at high doping, with a regime of
strong scattering in between, suggestive of a quantum phase transition
\cite{haule07,jarrell08b}. These results bear remarkable resemblance to those from
photoemission experiments on actual cuprates.
Because of the similarity to orbital-selective Mott physics, the partial disappearance of
the Fermi surface upon decreasing the doping has been termed ``momentum-selective Mott
transition''.\footnote{ In cluster DMFT, the momentum-selective Mott transition happens
via inter-cluster singlet formation which wins over the Fermi-liquid-like screening of
the cluster spins by the bath. This is exactly what happens in the cluster DMFT
description of Kondo-breakdown physics. }

The numerical data lend themselves to the following speculation: a non-Fermi liquid phase might
be realized at low doping, which is separated by a QPT from a Fermi liquid at large
doping. The non-Fermi liquid is characterized by small pockets of Fermi-liquid-like
carriers, which coexist with magnetic correlations carried by spin-1/2 local moments. The
QPT, likely located close to optimal doping, is accompanied by a critical Fermi surface
and could be responsible for broad quasiparticle spectra above the superconducting $\Tc$. As the
QCP is masked by superconductivity, these singularities are cut-off below $\Tc$, i.e.,
superconductivity is accompanied by the onset of coherence.

In this scenario, two questions are pertinent: Can this QCP be held responsible for the
unsually large values of $\Tc$ in cuprates? What is the true character of the low-doping
phase? The second question is particularly pressing, as a simple decoupling of a
Fermi-liquid-like conduction band and a local-moment spin liquid, as in the FL$^\ast$
phase of a two-band model, is not straightforward in the single-band case.
In a single-band description, it is unclear at present how fractionalization could be
compatible with experimental data (e.g. quantum oscillations) suggesting the existence of
Fermi-liquid-like quasiparticles.
In addition to further experiments, a more detailed analysis of numerical results might
help in guiding phenomenological ideas.

\subsection{Iron pnictide superconductors}

The iron pnictide superconductors have been on stage since 2008 \cite{kami08,rotter08},
and parallels to the
cuprates have been invoked frequently. By now, it seems clear that correlation effect are
significantly weaker and an ``itinerant'' description carries much further in the
pnictides as compared to the cuprates. In addition, orbital degrees of freedom play a
vital role in the physics of the pnictides, while single-band models are often assumed to
describe cuprates. Nevertheless, there are a number of properties of the pnictides which
may not be easily consistent with a multi-orbital weak-coupling picture, like the
bad-metal behavior typically found above the Neel temperature, the
significantly reduced bandwidth as compared to results from ab-initio calculations, and
the spectral-weight transfer in optical conductivity.

This has in turn stimulated approaches which assume correlation effects to be crucial
\cite{si08,phillips}. As parent compounds of the iron pnictides are not Mott-insulating,
those materials could be located on the metallic side in the vicinity to a Mott
transition. Considering the multi-orbital nature, an appealing idea is that they are
in fact near (or even inside, depending on doping or pressure) an orbital-selective Mott
phase \cite{phillips,kou,hackl09}. Whether such a picture is warranted could in principle
be deduced from ab-initio calculations, however, no generally accepted picture has
emerged here. While arguments have been put forward in favor of different correlation
strengths in different Fe 3d orbitals \cite{phillips}, this appears not supported by LDA
results \cite{mazinpriv,medici09b}. However, it has been shown that the lifting of
orbital degeneracy can be sufficient to drive an OS Mott transition \cite{medici09}.

Considering recent results from LDA+DMFT methods \cite{liebsch10a,liebsch10b,kotliar10},
a plausible scenario is that iron pnictides are moderately to strongly correlated
materials, where nearby Mott phenomena influence a variety of observables, but
which are not in a Mott or OS Mott phase. However, for somewhat larger correlation
strength (e.g. through chemical variations), such phases may be reached,
and here selective Mott transitions appear likely \cite{medici09b}.
In fact, orbital-selective Mott phenomena have been proposed for $\alpha$-FeSe on the
basis of ab-initio calculations \cite{craco09}.


\section{Conclusions}

In this article, we have reviewed the current theoretical understanding of
orbital-selective Mott phases and associated quantum phase transitions, which have
emerged in the fields of heavy fermion metals and multi-orbital transition metal oxides.
The zero-temperature orbital-selective Mott transition is a prime example of a
topological quantum phase transition, without a local order parameter. Consequently, it
cannot be described by the familiar Landau-Ginzburg-Wilson approach to quantum
criticality. Further fascinating aspects are critical Fermi surfaces, which naturally
accompany orbital-selective Mott transition, and the emergence of a stable non-Fermi
liquid phase in the orbital-selective Mott regime. We have also listed a number of
candidate materials where signatures of orbital-selective Mott physics have been
observed. It is clear that this field is a young one, with still limited solid knowledge,
and there is more exciting progress to come.


\begin{acknowledgements}
The author acknowledges collaborations with A. Hackl, S. Sachdev, and T. Senthil as well
as many helpful discussions with A. Benlagra, M. Brando, M. Civelli, P. Coleman, S.
Friedemann, M. Garst, P. Gegenwart, A. Hackl, G. Kotliar, L. de Medici, L. de Leo, H. v.
L\"ohneysen, I. Mazin, C. P\'epin, A. Rosch, S. Sachdev, T. Senthil, Q. Si, F. Steglich,
and P. W\"olfle.
This research was supported by the DFG through SFB 608, SFB-TR 12 and FG 960.
\end{acknowledgements}



\bibliographystyle{science2}

\end{document}